# Life & Light: Exotic Photosynthesis in Binary and Multiple Star Systems

J. T. O'Malley-James

School of Physics and Astronomy, University of St Andrews, North Haugh, St Andrews, Fife, KY16 9SS, UK

J. A. Raven

Division of Plant Sciences, University of Dundee at TJHI, The James Hutton Institute, Invergowrie, Dundee DD2 5DA, UK

C. S. Cockell

UK Centre for Astrobiology, School of Physics and Astronomy, James Clerk Maxwell Building, The King's Buildings, University of Edinburgh, Edinburgh, EH9 3JZ, UK

J. S. Greaves

School of Physics and Astronomy, University of St Andrews, North Haugh, St Andrews, Fife, KY16 9SS, UK

Corresponding Author:

J. T. O'Malley-James, School of Physics and Astronomy, University of St Andrews, North Haugh, St Andrews, Fife, KY16 9SS, UK, email: jto5@st-andrews.ac.uk

## ABSTRACT

The potential for hosting photosynthetic life on Earth-like planets within binary/multiple stellar systems was evaluated by modelling the levels of photosynthetically active radiation (PAR) such planets receive. Combinations of M and G stars in: (i) close-binary systems; (ii) wide-binary systems and (iii) three-star systems were investigated and a range of stable radiation environments found to be possible. These environmental conditions allow for the possibility of familiar, but also more exotic forms of photosynthetic life, such as infrared photosynthesisers and organisms specialised for specific spectral niches.

**INTRODUCTION**

One of the promising biosignatures of extraterrestrial life would come from life dependent upon oxygenic photosynthesis, exploiting its star as its primary energy source and producing the waste product, oxygen (Wolstencroft & Raven, 2002; Raven & Cockell, 2006), the spectroscopic detection of which in a terrestrial planet's atmosphere has been suggested as an indicator of the presence of life (Lovelock & Kaplan, 1975; Owen, 1980, Léger *et al*., 1993; Wolstencroft & Raven, 2002; DesMarais *et al*., 2002; Seager *et al*., 2005; Kaltenegger & Selsis, 2007). Oxygen could also be detected indirectly, for instance via the 9.6 µm ozone line (Angel *et al.*, 1986; Burke, 1986). Anoxygenic (non-oxygenating) photosynthesis is a likely evolutionary forerunner of oxygenic photosynthesis; therefore, if anoxygenic photosynthesis were present on an Earth-like planet (ELP), it is probable that biogeochemical changes would lead to an $O_2$ producing version (Raven, 2007; Raven 2009a,b).

Extra-solar planets of a similar mass to Earth have yet to be detected; however, exoplanets with masses $\leq 10$ $M_\oplus$ (where 1 $M_\oplus$ is the mass of Earth) have been confirmed. The recently published list of Kepler exoplanet candidates suggests that 2-4 $R_\oplus$ (where 1 $R_\oplus$ is the radius of Earth) planets are more abundant around cool stars (3600-4100 K) (Howard *et al*., 2011) and that Earth-mass planets may be common in the galaxy (Borucki *et al.*, 2011a,b). Estimates suggest approximately 18% of stars host a $< 10$ $M_\oplus$ planet (Wittenmyer *et al*., 2011) and Cantanzarite & Shao (2011) suggest that 1.4-2.7% of sun-like stars could host an Earth analogue planet. When Earth-sized planets are discovered and their spectra analysed, it will be possible to estimate the atmospheric composition and, crucially, detect the presence of liquid water (Scalo *et al.*, 2007; Jones 2008), a prerequisite for life as we know it (Hanslmeier, 2010). Such detection methods have already been tested using light curves and disk-averaged spectra of Earth, with studies concluding that spectral detection of the atmospheric signatures of photosynthesis and of Earth-like land-based vegetation (which exhibits a "red edge" in Earth's disk-averaged spectrum) on an ELP would be possible (Arnold *et al.*, 2002; Tinetti *et al.*, 2006; Kaltenegger *et al.*, 2007; Briot, 2009; Pallé, 2010).

Crucial for oxygenic photosynthesis is an appropriate atmosphere and sufficient light in the appropriate wavelength range. Life on Earth uses wavelengths from just below 400 nm to about 700 nm (rarely 730 nm) with the energy from two photons used in the transfer of an electron from $H_2O$ to $CO_2$. Photon availability for photosynthesis is determined by the emission properties of the Sun and attenuation by the atmosphere and (for aquatic organisms) natural waters (Wolstencroft & Raven, 2002; Kiang *et al.*, 2007a,b; Björn *et al.*, 2009; Milo, 2009; Raven, 2009a). Earth-analogue planets around stars for which the maximum photon emission is at different wavelengths than our Sun may result in oxygenic photosynthesis operating using a different number of photons to transfer an electron from $H_2O$ to $CO_2$; for example, photochemistry could theoretically operate with three photons of up to 1050 nm - 1100 nm (Wolstencroft & Raven, 2002; Kiang *et al.*, 2007b).

The possibility of habitable planets in binary star systems is not a new one (Harrington, 1977; 1981; Harrington & Harrington, 1978; Hale, 2006), but the influence of a second star on the potential for oxygenic photosynthesis has yet to be explored (the possibility for photosynthetic life in single star systems has been investigated in previous work (Franck *et al.,* 2001; Wolstencroft & Raven, 2002; Raven & Cockell, 2006; Cockell *et al*., 2009)). Extra-solar planets have now been found in binary and multiple stellar systems; most recently the Saturn-like planet Kepler 16b, which follows a circumbinary orbit, orbiting both stars in the system (Doyle *et al.*, 2011). To date, two planets below 10 $M_\oplus$: 55 Cnc e (a binary star system consisting of a G and M star separated by 1000 AU) (Fischer *et*

*al., 2008*) and GJ 667C b (a three-star system composed of a M dwarf orbiting a pair of K stars at 56-213 AU) (ESO, 2009). Of particular interest in this investigation are G stars (specifically G II V stars similar to the Sun) and M stars (low mass main sequence stars) – for the particular stellar parameters chosen for each star type used in this investigation see *Table 1*. The former are known to host a number of exoplanets and the latter are the most abundant stars in the galaxy (Tarter *et al.*, 2007; Guinan & Engle, 2009), with previous work showing that photosynthesis is theoretically possible on M star planets (Wolstencroft and Raven 2002; Kiang et al. 2007b). *Figure 1* shows that the type of systems of interest make up a major fraction of the binary systems surveyed by Duquennoy & Mayor, (1991) in a survey of 164 Sun-like primaries within the solar neighbourhood (within 250 ly). Approximately 57% of G stars are found in multiple star systems while M stars are less commonly found in multiple systems, having a multiplicity fraction of 25-30% (Lada, 2006; Tarter *et al.*, 2007; RECONS). While K stars would also be suitable for this investigation, including them would not add significantly to the conclusions formed here due to their similar properties to G stars, whereas, the order of magnitude difference in luminosity between G and M stars allows a wide range of potential flux densities to be explored.

Three scenarios involving combinations of high-mass (1 $M_\odot$) and low-mass (< 0.5 $M_\odot$) stars are explored: (i) Close binary systems (< 0.5 AU star-separation) with planets in p-type planetary orbits (in which the planet orbits both stars as if they were a single star), (ii) wide binary systems (> 3 AU star-separation) with s-type planetary orbits (in which the planet orbits just one of the stars) and (iii) multiple star systems consisting of two close stars and one more distant star (essentially a combination of cases (i) and (ii)).

## METHODS

The continuously habitable zone (CHZ), the region of space around a star(s) that remains habitable for long enough for life to emerge, is calculated following the method of Kolena (2007) by finding the distances required to maintain surface temperatures of 273 K and 373 K over the course of the host stars' stellar evolution. The calculation was performed using both the full fraction of the G star luminosity and 0.7 of that luminosity (to account for the increase in luminosity over a main sequence G star's lifetime, with 0.7 $L_\odot$ being the luminosity of the Sun 4.5 Gyr ago predicted by stellar evolution theory) to find the region that remains habitable for approximately 4.5 Gyr (M star luminosities change very little over such time periods (Tarter *et al.*, 2007) so only one calculation was necessary in those cases). Although it is not certain that the early Sun had a luminosity of 0.7 $L_\odot$ (Gaidos *et al.*, 2000), it is convenient for the purposes of this investigation to assume this value.

A plethora of factors influence the habitability of a planet. Finding a planet within the CHZ does not necessarily imply that conditions will be conducive to life. While it is impractical to factor all these components into this investigation, one important factor to consider is the stability of planetary orbits in these binary system scenarios.

The region in which stable orbits are possible for both p-type orbits (orbits in which the planet orbits both planets in the binary system) and s-type orbits (orbits in which the planet orbits one of the two stars in the system) in binary systems can be estimated using the boundary conditions given in Holman & Wiegert (1999) where mass, eccentricity and star separation data for test particle simulations were used to form empirical expressions for the critical orbital distance of a planet in both p-type and s-type scenarios. Given a mass ratio (μ) and star separation (a), and assuming an eccentricity (e) of zero, these boundary conditions serve as guidelines for how realistic placing a planet in the CHZ in these systems would be. Under these conditions the equations given for maximum and minimum distance for stable orbits reduce to

$$a_p = [1.6 + 4.12\mu + (-5.09\mu^2)]a \qquad (1)$$

for p-type planetary orbits and

$$a_p = [0.464 + (-0.38\mu)]a \qquad (2)$$

for s-type orbits. In *Equation 1*, $a_p$ expresses a minimum distance because, for a p-type orbit, getting too close to the two stars is the most limiting factor to a stable orbit. Conversely, a maximum distance is given by *Equation 2*, because, a planet would cease to be in an s-type orbit if it moved too far away from its host star that the gravitational attraction of the second star began influencing the orbit. In order to ensure that planets fall within the CHZ, $a_p$ is fixed to a distance within the CHZ in each case and the appropriate value for the star separation, $a$ is calculated.

Low eccentricities improve the probability that a planet will remain habitable for (geologically) long time spans by reducing extreme environmental fluctuations (Kita *et al.*, 2010). While Dressing *et al.* (2010) suggest that ELPs with eccentricities as high as 0.9 may still be habitable, such planets would experience more extreme environmental variability, making them less desirable targets. From looking at known exoplanet eccentricities (*Figure 2*), there is a reasonable fraction of low-mass planets that have low eccentricities, suggesting it is not unreasonable to assume this could be the case for ELPs. Of the known terrestrial planets in multiple star systems, Cnc e has an eccentricity of 0.07, while the eccentricity of GJ 667 C b is currently not known.

A planet in a stable orbit within the CHZ can then be assessed in terms of its suitability for oxygenic photosynthesis. On Earth, oxygenic photosynthesis uses a range of wavelengths that are constrained by a number of environmental, cell energetic and evolutionary factors (Wolstencroft & Raven, 2002; Falkowski & Raven, 2007; Kiang *et al.*, 2007a,b; Stomp *et al.*, 2007; Björn *et al.*, 2009; Milo, 2009; Raven, 2009a).

The environmental factors are the range of wavelengths over which the Sun provides a high photon flux density, as modified by attenuation by the atmosphere and, for aquatic organisms, by natural waters. The need to avoid damaging UV radiation restricts the use of shorter wavelengths and the energetics of photochemical reactions restricts longer wavelengths. The wavelength range used in the photochemical reactions of photosynthesis is centred on wavelengths (680 nm, 700 nm) just above that of maximum photon flux from the Sun (Wolstencroft & Raven, 2002; Raven, 2011); hence, knowing the maximum photon flux from stars allows assumptions to be made about the probable reaction centre wavelength used in photosynthesis. The range of wavelengths over which photons are harvested is about 300 nm on the short wavelength side of the wavelengths used in photochemistry.

By first determining the wavelength of maximum photon flux ($\lambda_{max}$) using Wien's Law in its photon basis (Soffer & Lynch, 1999; Nobel, 2005) for the star(s) in a given scenario, the photon flux density (PFD) in µ mol photons m$^{-2}$ s$^{-1}$ on which photosynthesis depends can be found using

$$PFD = \left[\frac{1}{4}\left(\frac{L_*}{4\pi d^2}\right)\left(\frac{1}{hc/\lambda_{max}}\right)\right]\left(\frac{1\times 10^6}{N_A}\right) \qquad (3)$$

where $L_*$ is the luminosity associated with the star, $d$ is the mean distance between the planet and star and the $10^6/N_A$ term (where $N_A$ is the Avogadro constant) ensures the flux density is counted in µ mol photons (McDonald, 2003; Puxley *et al.*, 2008).

It is also useful to know a planet's likely orbital period. For the wide binary cases, radiation from each star hits an area totalling half of the planet's surface. These areas can be completely separate or overlap partially or completely. When there is no overlap the entire planet is illuminated, half by one star and half by the other. Conversely, when there is complete overlap, only half of the planet is illuminated, intercepting radiation from both stars. These two cases are half an orbit apart, hence, knowing the orbital period allows the duration of these different radiation regimes to be estimated.

For the purpose of this analysis it is assumed that all other conditions for photosynthetic life are met, for example a sufficient partial pressure of $CO_2$ and sufficient nutrients to support the organisms (Franck *et al*., 2001; von Bloh *et al*., 2010).

**RESULTS**

The stellar parameters used in these calculations are summarised in *Table 1*. Using equations *(1)* and *(2)* it was found that all scenarios permitted stable orbits within the CHZ given reasonable star separations (see *Table 2*).

The calculations of $\lambda_{max}$ for G and M stars resulted in peaks at 643.9 nm and 991.9 nm respectively. This is consistent with results of similar calculations, falling within the range of values quoted in recent literature (Wolstencroft & Raven, 2002; Nobel, 2005; Kiang *et al*., 2007b; Milo, 2009); the conclusions drawn here would not be significantly altered by using other values within this range.

For close-binary configurations, M-M star, M-G star and G-G star combinations were considered in orbits close enough that the system dynamics approximated a single star. Similar cases are investigated for wide-binary configurations, only a distinction is drawn between which star hosts the planet. In the trinary star cases, combinations of these wide- and close-binary scenarios are used. The results are presented in *Figure 3*.

In each case the stars were assumed to be separated by the minimum distance for stable planetary orbits within the CHZ in order to best investigate the influence of the two different radiation regimes. The planet was given a semi-major axis ($a_p$) that placed it within the CHZ such that the photon flux from both stars was maximised. An orbital eccentricity of zero was assumed.

# DISCUSSION

For star systems with a combination of two, or more, types of star, the light exploited by photosynthetic life would depend upon the arrangement of the star system in question (Cockell *et al.*, 2009). A planet orbiting close enough to an M star with a more distant G star companion would have a photosynthetic light regimen dominated by infrared radiation. The available PFD would be more than adequate to compensate for the theoretical need for a greater quantity of photons for oxygenic photosynthesis at such high wavelengths due to the close-in CHZ. Therefore, any life on such an ELP would likely show similar characteristics to those required in the single M star case.

**Infrared Photosynthesis in M star radiation regimes**

Infrared photosynthesis is known in anoxygenic photosynthetic organisms on Earth (Heath *et al.*, 1999) and it is theoretically possible that infrared photons could power oxygenic photosynthesis using more than two photons to move an electron from water to carbon dioxide (Heath *et al.,* 1999; Wolstencroft & Raven, 2002; Raven 2009a). The anoxygenic photosynthetic bacteria harvest photons for photosynthesis in the visible and infrared using chlorophyll derivatives that, in association with the appropriate protein, can absorb at wavelengths of 400 nm to, in some cases, almost 1000 nm; most of these organisms cannot carry out photochemistry in the presence of oxygen (Kiang *et al.*, 2007a; Stomp *et al.* 2007; Raven 2009b). Certainly, the concept of spectral niches for photosynthetic life (Raven, 2007; Björn *et al.*, 2009), based on depth or position in an aquatic environment that this example highlights would be equally valid on other ELPs. However, on Earth, none of these organisms evolve oxygen (Kiang *et al.*, 2007a; Raven 2009b), and those that only carry out photosynthesis in anoxic conditions use other reductants, producing by-products such as sulphur, the occurrence of which could be explained by non-biological processes. While it may be possible to detect sulphuric acid biosignatures, there would be the risk of false-positive results (Domagal-Goldman *et al.*, 2011). On Earth there are also bacteria which carry out photochemistry under aerobic conditions to economise on the use of organic carbon by partly replacing ATP from respiration with photochemically produced ATP, without leaving an extracellular oxidised inorganic product or reducing $CO_2$ (Gasol *et al.* 2008, Raven 2009b). Thus, anoxygenic photosynthesis, if present, would be more difficult to detect remotely.

Due to the close-in habitable zone in M star systems, a protective mechanism against UV flares would be necessary, especially as flare activity may take longer to die down in binary systems due to the interactions between the stars (Heath *et al.*, 1999). Protection against the rapid change in UV flux associated with solar flares may be an important issue for life, and especially photosynthetic life, which must be exposed (directly or indirectly) to stellar radiation on ELPs within such systems. Photosynthetic organisms on Earth have a range of screening and motility mechanisms that restrict exposure to UV radiation relative to longer wavelengths, or repair damage, in cyanobacteria (Garcia-Pichel & Castenholz, 1993), eukaryotic photosynthetic organisms (Garcia-Pichel & Castenholz, 1993; Oren & Grunde-Cimerman, 2007) and embryophytic plants (Heath *et al.*, 1999; p.273 of Inderjit *et al.*, 1999; Turunen & Latola, 2005). UV radiation flares may also be avoided by photosynthetic organisms within substrates such as rock and salt or underwater (Cockell, 1999; Cockell *et al.*, 2009). UV protection could also come from thick cloud cover on a planet (Mayer *et al.*, 1998).

The possibility of photosynthetic machinery on ELPs that uses UV radiation cannot be discounted, since UV-A radiation can be used to power photosynthesis, at least over short time intervals (Haldall, 1964, 1967; Kawaguti, 1969; Neori *et al.*, 1986, 1988; Schlichter *et al.*, 1996; Gao *et al.*, 2007; cf. Bühlman *et al.*, 1987; Salih *et al.*, 2000). Starspots may also influence the nature of life on ELPs, being proportionally larger than those of our own Sun, causing a 10-40% decrease in insolation (Heath *et al.*, 1999) compared to the few tenths of a percent decrease caused by solar sunspots (Fröhlich & Lean, 2004). Therefore it is likely that native organisms would be adapted to the lower temperature periods associated with starspot activity.

**Dominant Vegetation Colour**

The spectral range of photosynthetic pigments on a planet is likely to differ depending on the type of star that planet orbits (Wolstencroft and Raven 2002; Kiang *et al.* 2007a; 2007b; 2008). Hence, vegetation adapted to an M star environment would likely appear a different colour to that on Earth (although the concept of "colour" can become more complicated when absorption/reflection/transmission are not all in our visible range). As pointed out by Wolstencroft & Raven (2002) a longer wavelength limit would be expected for absorption by photosynthetic pigments, with the energy of photons used in photochemistry corresponding to the long wavelength limit (Björn *et al.*, 2009; Milo, 2009; Raven, 2009a). At the short wavelength limit, black body radiation would have relatively little energy in the blue region where there are major absorption bands of chlorophylls, and absorption by carotenoids.

Cost-benefit criteria would apply to the wavelength, and quantity, of any additional antenna pigments used by the organisms in an M star environment to those on Earth (Raven, 1984), and it would not be expected that total (black body) absorption of photons occur in the M star environment any more than occurs on Earth. However, in M star radiation habitats, vegetation may have more kinds of photosynthetic pigments in order to make use of a fuller range of wavelengths. Some plants on Earth are almost equally absorptive at all wavelengths in the part of the spectrum being considered, but are reflective in the UV band (a mechanism evolved in relation to attracting insects which can see UV radiation) (Chittka *et al.*, 1994; Kevan *et al.*, 2001). Therefore, the ability to reflect UV radiation to a certain extent could help to avoid its biologically damaging effects.

**Spectral Niche Variation in Multiple Radiation Regimes**

Another question is how photosynthetic life might be adapted to the presence of two different radiation environments: would it preferentially use G star radiation, M star radiation, or a combination of both?

For the example of an M and G star in a close binary arrangement, a habitable planet would be approximately 1 AU from the barycentre. At this distance the photon flux density from the M star would be much lower than that from the G star, suggesting that G star radiation would be preferred. While there may be some environments on such a habitable ELP that make use of infrared photons preferable, perhaps due to the light attenuation properties of a certain habitat, it must be borne in mind the maximal use of G star radiation in oxygenic photosynthesis involves the use of two photoreactions (at a wavelength of 700 nm) to move each electron from water to carbon dioxide, while using M star radiation would require more than two reactions at longer wavelengths (Wolstencroft & Raven, 2002). Similar arguments apply to the G-M wide binary, only now the flux from the more distant M star is much less likely to be useful for photosynthesis. The M-G wide binary presents the opposite case where radiation from the M star is more likely to be used, but the use of G star radiation may be preferred in some circumstances.

The case for photosynthetic organisms adapted to use both forms of radiation is harder to make. It would be complicated and expensive in terms of energetic investments to house both of these systems in a single organism. While two oxygenic phototrophs are known to have rhodopsin-based as well as chlorophyll-based photochemical energy conversion, the rhodopsin-based system has pigments which absorb within the same wavelength range as the chlorophyll-based system, and the rhodopsin-based system could only be an auxiliary component of oxygenic photosynthesis (Raven, 2009a).

The possibilities become more interesting for the wide binary M-G star scenarios. The primary star's radiation always has a greater magnitude than that of the distant secondary star (see *Figure 3*); however, there are periods where a portion of the planet would be illuminated only by light from the less photosynthetically favourable secondary star. When the G star is the planet-hosting star, the M star is too distant to make any useful contribution to a habitable planet's photosynthetically active radiation (PAR). However, if the M star hosts a planet, some organisms may evolve to exploit the low photon flux density from the distant G star, as G-star-only illumination can persist in some regions on the planet's surface for a significant portion of its orbit. On Earth, when PAR is unavailable for an extended period of time, most photosynthetic organisms can survive by using major macromolecular pools as shown for the cyanobacterium *Phormidium autumnale*, which is able to survive for at least three weeks in continuous darkness without using any external organic matter (Montechario *et al.*, 2006; Montechario & Giordano, 2006). In this case the exact energy reserve compound was not identifiable because the organism consumed carbohydrate, protein and lipids in proportional quantities, rapidly resuming photosynthesis and growth upon re-exposure to PAR (Montechario *et al.*, 2006; Montechario & Giordano, 2006). It should be noted that the presence of the secondary G star moves the CHZ further away from the M star (allowing an Earth-analogue planet to exist comfortably at 1 AU from the M star given a star separation of 1.5 AU), reducing the risks associated with UV flares.

On Earth specialists tend to be more successful than generalists when there is a large range of resources available in a habitat (Ferry-Graham *et al*., 2002). The changing radiation regimen in M-G binary systems raises the possibility of spectral niche variation, with different organisms coexisting in the same habitat adapted to use available radiation from different stars at different times in the planet's orbit. Evolving a mechanism in a single organism to exploit both types of radiation available in a habitat, switching between the two in response to the changing environment may be too costly a strategy to be favourable. It is more likely that two distinct categories of organism would evolve under these conditions; each adapted to make use of one form of radiation or the other. However, there are examples among photosynthetic algal flagellates on Earth in which oxygenic photosynthesis co-exists with the capacity to ingest smaller organisms (phagotrophy) as a means, complementary to photosynthesis, of obtaining organic carbon (and other nutrients); such mixotrophy is common among marine flagellates (Gasol *et al.* 2008; Raven *et al.* 2009).

**Triple Systems**

The models investigated suggest that the presence of a third star does not significantly influence the radiation environment on a planet within that system. It either approximates that of a single star system, or a close binary system, depending on the orbit the planet follows within the system. Similar arguments as above would apply to any photosynthetic life within such systems. GJ 667 is a triple system that is known to host a planet and there are many triple, or multiple stellar systems in the galaxy (13 systems were found in the Duquennoy & Mayor (1991) survey) to make these valid targets in the search for life.

## CONCLUSIONS

There are a number of different arrangements of multiple star systems in which ELPs can be classed as habitable; each providing unique environments and a variety of possibilities for oxygenic photosynthetic life. Changing spectral regimens in such systems would shift the dominantly available radiation between infrared and red radiation at different points in a planet's orbit, raising the possibility of organisms with two different forms of photosynthetic machinery, or two distinct biospheres, each adapted to use the specific radiation associated with one of its suns. The results presented here are sufficient to demonstrate that binary and multiple star systems are plausible targets in the search for extrasolar oxygenic photosynthesis.

## ACKNOWLEDGEMENTS

The authors wish to thank Dr Carsten Weidner for helpful discussions on orbital stability and four anonymous reviewers for comments that significantly improved the manuscript. This paper is derived from a dissertation submitted by Jack O'Malley-James in partial fulfilment of the requirements for an MRes in Environmental Biology at the University of St Andrews. The University of Dundee is a registered Scottish charity, No. SC05194.

## AUTHOR DISCLOSURE STATEMENT

No competing financial interests exist.

**TABLES**

| Star | G Star | M Star |
|---|---|---|
| **Mass, $M_\odot$** | 1 | 0.3 |
| **Radius, $R_\odot$** | 1 | 0.4 |
| **Luminosity, $L_\odot$** | 1 | 0.06 |
| **Temperature, K** | 5700 | 3700 |
| **Photon wavelength at peak flux density, nm** | 643.9 | 991.9 |

**Table 1:** Stellar Properties: Values used for G and M star properties in this investigation (Wolstencroft & Raven, 2002).

| Scenario | CHZ (AU) | Star Separation for Orbital Stability in CHZ, a (AU) | $a_p$ (AU) | PFD (μ mol photons m$^{-2}$s$^{-1}$) | Temperature Contribution (K) |
|---|---|---|---|---|---|
| M-M [p] | 0.2-0.4 | 0.1 | 0.3 | 2000 (M) <br> 2000 (M) | 273 (combined) |
| M-G [p] | 0.6-1.0 | 0.5 | 0.6 | 5000 (G) <br> 500 (M) | 333 (combined) |
| G-G [p] | 0.9-1.4 | 0.5 | 0.9 | 2250 (G) <br> 2250 (G) | 318 (combined) |
| M-M [s] | 0.2-0.3 | 2.0 | 0.2 | 4200 (M) <br> 42 (M) | 281 (M) <br> 89 (M) |
| G-M [s] | 0.6-1.0 | 4.0 | 1.0 | 2000 (G) <br> 10 (M) | 254 (G) <br> 62 (M) |
| M-G [s] | 0.2-0.3 | 2.0 | 0.5 | 675 (M) <br> 456 (G) | 178 (M) <br> 179 (G) |
| G-G [s] | 0.6-1.0 | 4.5 | 1.0 | 2000 (G) <br> 90 (G) | 254 (G) <br> 119 (G) |
| (M-M)-G (i) | 0.7-1.0 | 3.0 | 0.85 | 230 (M) <br> 230 (M) <br> 200 (G) | 162 (M-M) <br><br> 146 (G) |
| M-M-(G) (ii) | 0.6-1.0 | 3.0 | 1.0 | 2000 (G) <br> 20 (M) <br> 20 (M) | 254 (G) <br> 86 (M-M) |
| (G-G)-G (i) | >2.0 | > 3.0 | 2.0 | 500 (G) <br> 500 (G) <br> 200 (G) | 213 (G-G) <br><br> 146 (G) |
| (G)-G-G (ii) | >2.5 | > 3.0 | 2.5 | 300 (G) <br> 200 (G) <br> 200 (G) | 160 (G) <br> 174 (G-G) |

**Table 2:** Summary of orbital parameters, mean annual photon flux densities and individual temperature contributions for the multi-star scenarios investigated. The symbols [p] and [s] denote p-type and s-type planetary orbits respectively. G and M denote G stars and M stars respectively. For s-type orbits, the first letter in the scenario description denotes the star which the planet orbits. For the triple star systems, the brackets denote the star (or component) orbited by the planet.

**FIGURE LEGENDS**

**Figure 1:** Proportion of G-M star and G-G star binaries (and multiple star systems) in the sample of systems with G star primaries from Duquennoy & Mayor, 1991. The type of system of interest in this work appears to be frequent enough to warrant investigation.

**Figure 2:** Eccentricity values for known exoplanets in single and multiple G and M star systems. Source: The Extrasolar Planets Encyclopaedia.

**Figure 3:** *Peak (Photosynthetic) Photon Flux Density Results:* Red denotes the PFD from the M stars and yellow the PFD from the G stars in any given configuration. The lower red line indicates the peak photon flux density on Earth and the upper red line indicates the theoretical photon flux density required for infrared photosynthesis such that photosynthetic productivity equals that on Earth. *(1) Close binaries:* The close binary scenarios all approximate single star systems in terms of orbital dynamics and light regimes. G star radiation clearly dominates in the G-M star close binary example; *(2) Wide binaries:* The M-M, G-M and G-G cases are all clearly dominated by one particular radiation envrionment. In the M-G case, although the flux from the second star is low, it could still potentially be exploited for photosynthesis; *(3) Triple systems:* All but the M-M-G(ii) fall below the threshold for an Earth-like level of photosynthesis.

## G-star Binary Systems, Duquennoy & Mayor, 1991

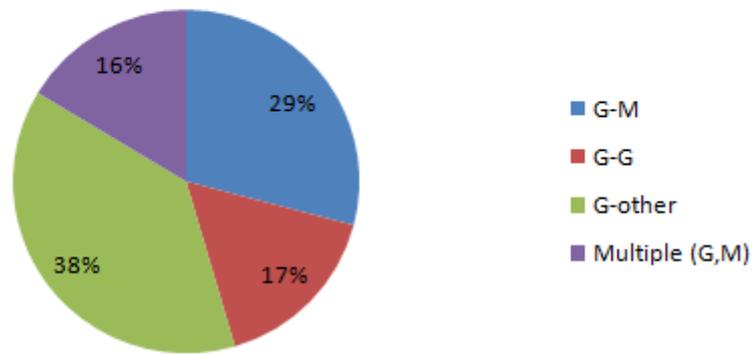

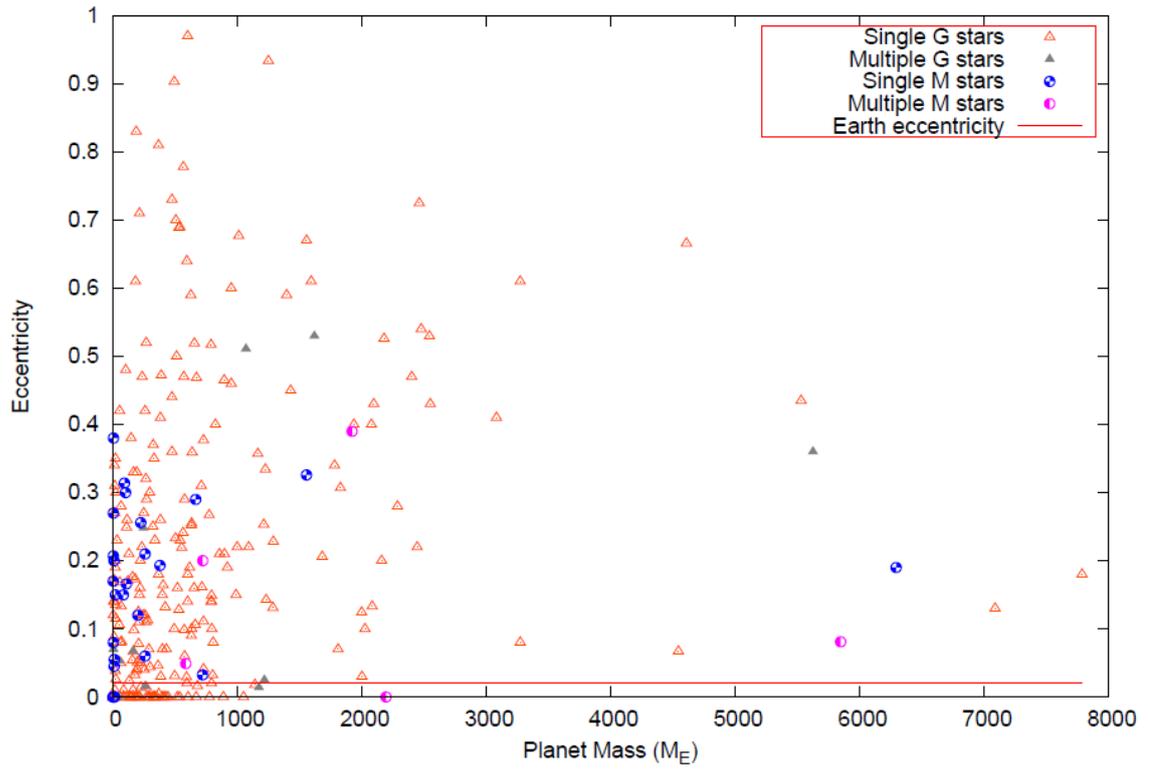

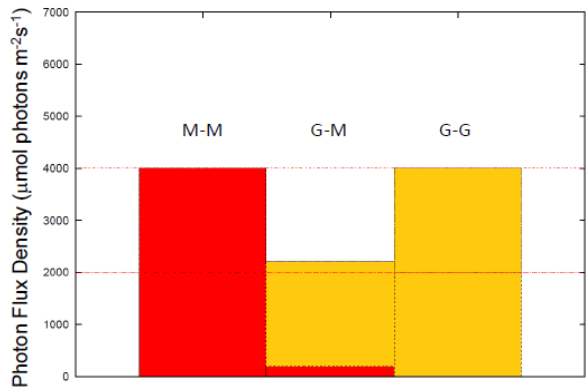
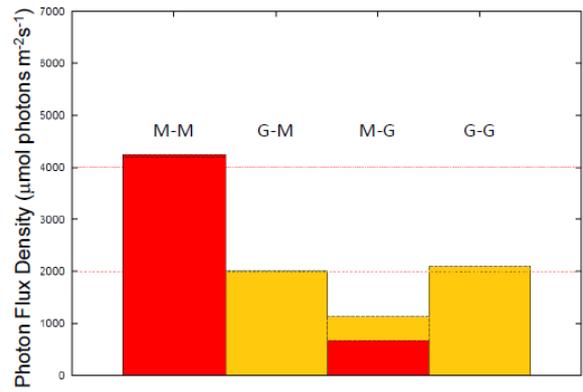
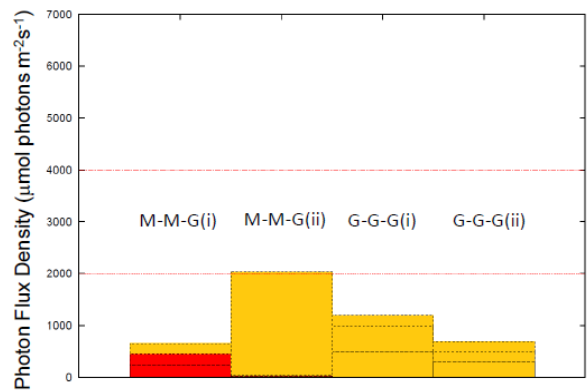